\documentclass[aps,twocolumn,floatfix,superscriptaddress,longbibliography,nofootinbib]{revtex4-1}
\pdfoutput=1

\usepackage{graphicx}
\usepackage{indentfirst}
\usepackage{latexsym}
\usepackage{multirow}
\usepackage{tabls}
\usepackage{epsfig}
\usepackage{multirow}
\usepackage{subfigure}
\usepackage{comment}
\usepackage{hyperref}
\usepackage{color}
\usepackage[usenames,dvipsnames]{xcolor}
\usepackage{amssymb}
\usepackage{amsfonts}
\usepackage{amsmath}
\usepackage{bm}
\usepackage{mathrsfs}
\usepackage{dsfont}

\begin{document}

\title{Quantifying and mitigating bias in inference on gravitational wave source populations}

\def\addCambridge{Institute of Astronomy, Madingley Road, Cambridge, CB30HA, United Kingdom}

\author{Jonathan R. Gair}
\email{jrg23@ast.cam.ac.uk}
\affiliation{\addCambridge} 

\author{Christopher J. Moore}
\email{cjm96@ast.cam.ac.uk}
\affiliation{\addCambridge} 

\date\today

\begin{abstract}
When using incorrect or inaccurate signal models to perform parameter estimation on a gravitational wave signal, biased parameter estimates will in general be obtained. For a single event this bias may be consistent with the posterior, but when considering a population of events this bias becomes evident as a sag below the expected diagonal line of the P-P plot showing the fraction of signals found within a certain significance level versus that significance level. It would be hoped that recently proposed techniques for accounting for model uncertainties in parameter estimation would, to some extent, alleviate this problem. Here we demonstrate that this is indeed the case. We derive an analytic approximation to the P-P plot obtained when using an incorrect signal model to perform parameter estimation. This approximation is valid in the limit of high signal-to-noise ratio and nearly correct waveform models. We show how the P-P plot changes if a Gaussian process likelihood that allows for model errors is used to analyse the data. We demonstrate analytically and using numerical simulations that the bias is always reduced in this way. These results provide a way to quantify bias in inference on populations and demonstrate the importance of utilising methods to mitigate this bias.
\end{abstract}

\maketitle

\section{Introduction}
In the coming years it is expected that the advanced era ground-based gravitational wave (GW) detectors that are now coming online (such as advanced LIGO \cite{2010CQGra..27h4006H} and advanced Virgo \cite{Acernese2009}) will begin to make routine measurements of GWs from a variety of sources. Later in this decade, pulsar timing arrays could also begin to detect sources of nanohertz gravitational waves~\cite{NANOMaura:2013,Hobbs:2013CQG,KramerCha:2013CQG,Manchester:2013CQG} and there are ambitious plans for a space-based gravitational wave detector (eLISA~\cite{TheGravitationalUniverse}) operating in the millihertz band, that with be launched by ESA around 2034. Inferences about source parameters in this new era of GW astronomy will rely on the availability of detailed signal models for the sources. The calculation of accurate models is computationally prohibitive, however, so approximate models will be used for inference, which will, in general, lead to biases in the parameter estimates obtained. This can lead one to make incorrect inferences about individual sources as well as incorrect inferences about astronomical populations of sources. The bias due to incorrect models becomes more important for louder sources. eLISA  is expected to observe the inspiral and merger of supermassive black holes at signal-to-noise ratios of ${\cal{O}}(10^{3})$. The impact of parameter bias has been shown to be even more significant in this case \cite{CV}.

A common way to quantify the performance of parameter estimation is via the probability-probability (P-P) plot. The P-P plot shows the probability that the true source parameters will lie in a given confidence interval estimated from the detector data, against the value of the confidence interval. In the ideal, unbiased, case the P-P should be a diagonal line; i.e.,\ $x$\% of the time the true source parameters should lie with the $x$\% confidence interval. However, there are a variety of effects that can cause the P-P plot to deviate from this ideal. For example, use of a greedy algorithm to build a multi-dimensional confidence interval from a kD-tree constructed from a random sample of points from a distribution (this problem was discussed in the context of sky-localisation by \cite{2014PhRvD..89h4060S}), deviations between the waveform model and the true signal due to a breakdown of general relativity (GR) in the strong field (the case of undetectable deviations from GR, the so-called ``stealth-bias'', was considered in \cite{2013PhRvD..87j2002V}), and mis-estimating the noise properties of the detector can all cause the P-P plot to deviate from a ideal diagonal line. However, the cause of biased parameter estimation that we will consider in this paper is the presence of inaccuracies in the waveform model used to analyse the data~\cite{CV}. If such a systematic error is present the returned confidence intervals from a parameter estimation study will be shifted away from the true parameters making it less likely that the confidence interval contains the true parameters. Therefore the P-P plot will ``sag'' below the ideal diagonal line. 

Recently \citep{Moore&Gair} the authors proposed a marginalised likelihood which uses Gaussian processes (GPs) to fold in extra information from a small \emph{training set} of accurate waveforms, e.g.\ numerical relativity (NR) waveforms. Accurate here refers to how well these waveforms represent solutions of the GR field equations. Numerical relativity waveforms are not perfectly accurate, but they are the best solutions currently available and inaccuracies in them can be folded into the GP analysis. If astrophysical gravitational waves are governed by a theory other than general relativity, these waveforms will not be accurate representations of reality. This will also lead to a bias, but one that is harder to quantify without knowing the true theory of gravity. Here we proceed assuming GR is correct and look only at biases from model uncertainties. Once observations are made this assumption could be revisited if evidence arises for departures from GR.

The GP marginalised likelihood in general shifts the best fit parameters closer to the true parameters and broadens the peak in the posterior, making it more likely that a given confidence contour contains the true parameters. Therefore, it would be expected that parameter estimates obtained using the marginalised likelihood would exhibit less of a bias, and the P-P plots would exhibit less of a ``sag''. However, the Gaussian process regression (GPR) which underlies the marginalised likelihood makes some assumptions about how the error in the waveform model varies over parameter space. In this paper, we investigate the P-P plots both in the case where these assumptions turn out to be correct, and, more importantly, when they are incorrect.

There are two main results in this paper. The first is a derivation of an analytic expression for the expected sag in a P-P plot arising from waveform uncertainties. This is derived under the assumption that the waveform error is small so that we can use the linear signal approximation. The second is that the use of the marginalised likelihood constructed via Gaussian process regression to analyse data leads to a reduction in the size of the deviation from the diagonal line. The sag is removed completely if the true waveform errors are drawn from the same model used to construct the marginalised likelihood. However, even when the errors follow a different distribution, the marginalised likelihood leads to a reduction in the sag.

This paper is organised as follows. Sec.\ \ref{sec:PE} provides a recap of and quotes some necessary results about GW parameter estimation, and introduces the marginalised likelihood. Sec.\ \ref{sec:deriv} derives analytic expressions for the P-P plots for both the standard and marginalised likelihoods for a variety of possible waveform errors. Sec.\ \ref{sec:numerics} describes the numerical simulations that were performed to back-up the analytic results in Sec.\ \ref{sec:deriv}. Finally Sec.\ \ref{sec:conclusions} contains a discussion of the results and concluding remarks.

\section{Parameter estimation}\label{sec:PE}
We assume that the source of GWs is fully specified by a parameter vector $\vec{\lambda}$, and that the true waveform model is $h(t;\vec{\lambda})$ (hereafter the dependence of $h$ on time $t$ is supressed for clarity). The aim of a parameter estimation study given measured data $s$, is to estimate the posterior probability on the parameters, $P(\vec{\lambda}|s)$. This is given from Bayes theorem (Eq.~\ref{eq:Bayes}) by the likelihood, $P(s|\vec{\lambda})\equiv L'(\vec{\lambda})$, the prior, $P(\vec{\lambda})$, and the normalising Bayesian evidence $Z=\int\textrm{d}\vec{\lambda}\,P(\vec{\lambda})L'(\vec{\lambda})$;
\begin{equation}\label{eq:Bayes} P(\vec{\lambda}|s)=\frac{P(\vec{\lambda})L'(\vec{\lambda})}{Z} \,.\end{equation}
As this paper concerns parameter estimation, and not model selection, we will not discuss the evidence further, since for any given source, this just enters as a normalisation factor for the posterior. In the case of stationary, Gaussian, additive noise $n$ in the detector the measured data is given by $s=h(\vec{\lambda}_{0})+n$ and the likelihood is given by
\begin{equation}\label{eq:Lexact} L'(\vec{\lambda})\propto\exp\left(-\frac{1}{2}\left<s-h(\vec{\lambda})\big|s-h(\vec{\lambda})\right>\right)\,, \end{equation}
Where $\left<\cdot | \cdot\right>$ denotes the usual noise-weighted inner product
\begin{equation}\label{eq:inner}\left<a \big| b \right> = \int_{-\infty} ^{\infty} \frac{ \tilde{a}^*(f) \tilde{b}(f)}{S_n(f)}  {\rm d} f \,.\end{equation}
In Eq.~(\ref{eq:inner}), $S_{n}(f)$ is the (two-sided) noise power spectral density in the detector.

In general we do not have access to the true waveform model $h(\vec{\lambda})$, at least not at a reasonable computational cost. Highly, but not totally, accurate NR waveforms have recently started to become available \citep{PhysRevLett.95.121101}, and slightly less accurate (but computationaly cheaper) extended analytic models such as (S)EOBNR \citep{2011PhRvD..84l4052P} are also available. However, these are too computationally expensive to use in routine parameter estimation studies, which typically require many thousands of likelihood evaluations. Instead, we must make use of cheaper but less accurate waveforms, such as post-Newtonian (PN) \citep{lrr-2014-2}), or numerical ``kludge'' models \citep{2007PhRvD..75b4005B}. Denoting the approximate waveform model by $H(\vec{\lambda})$, the {\it approximate likelihood} obtained when using this model is given by
\begin{equation}\label{eq:Lapprox} L(\vec{\lambda})\propto\exp\left(-\frac{1}{2}\left<s-H(\vec{\lambda})\big|s-H(\vec{\lambda})\right>\right)\,.\end{equation}
In general, posterior distributions obtained from this likelihood will not agree with posterior distributions obtained from the exact likelihood in Eq.~(\ref{eq:Lexact}). Denote by $\vec{\lambda}_{\textrm{exact}}$ the best fit parameters obtained from Eq.~(\ref{eq:Lexact}) and $\vec{\lambda}_{\textrm{approx}}$ the best fit parameters obtained from Eq.~(\ref{eq:Lapprox}). If both the waveform difference and the parameter shift $\Delta \vec{\lambda}\equiv \vec{\lambda}_{\textrm{approx}}-\vec{\lambda}_{\textrm{exact}}$ are small quantities, ${\cal{O}}(\epsilon)$, then an approximate expression for the shift in the parameters can be found by expanding in $\epsilon$. The shift in best-fit parameters to linear order in $\epsilon$ was obtained in \cite{CV} as $\Delta\vec{\lambda}\equiv\Delta \vec{\lambda}_{1}$ where
\begin{equation}
\Delta\lambda_{1}^{a} = - \left( \Sigma^{-1} \right)^{ab} \langle \delta h(\vec{\lambda}_{0}) | \partial_{b} H(\vec{\lambda}_{0}) \rangle \,,\label{CVsolmain}
\end{equation}
$\Sigma_{ab}=\left<\partial_{a} H(\vec{\lambda})|\partial_{b} H(\vec{\lambda})\right>$, and $\partial_{a}=\partial/\partial \lambda^{a}|_{\vec{\lambda}=\vec{\lambda}_{0}}$. For completeness we include a derivation of this result, and an extension of it to quadratic order, in Appendix~\ref{appendix}). 

From Eqs.~(\ref{CVsolmain}) and (\ref{secondorder}) it can be seen that the systematic shift in parameters caused by using the approximate likelihood is independent of the signal-to-noise ratio (SNR). This fact was observed in \cite{CV}, and since the statistical errors that arise from detector noise decrease with increasing SNR this means that the systematic shift is most important for the loudest sources.

When using the approximate likelihood in Eq.~(\ref{eq:Lapprox}) to characterise a single source one would usually use the condition that the systematic error due to the model uncertainty is less than the random error arising from noise to determine if the model is ``good enough''. This condition ensures that the true parameters will be consistent with the posterior --- the amount by which the systematic error shifts the peak of the posterior is less than the typical posterior width. However, whilst this condition ensures that the true parameters will always be consistent with the posterior, on average they will be further from the centre of the posterior and hence lie at a lower significance than they should. This starts to become important when observing a population of sources (as we hope will be the case for  Advanced LIGO). Even small systematic shifts may lead one to make incorrect inferences about the properties of the population. This can be understood by imagining that we observe a NS-NS binary with identical astrophysical parameters $n$ independent times with Advanced LIGO. The error in the combined estimate for the mean mass of the population is the error in each measurement divided by $\sqrt{n}$. Therefore even if the systematic model error is insignificant for making inferences regarding a single binary it becomes increasingly significant for inferences regarding populations as new sources are added. The importance of the model errors for LIGO observations of NS-NS binaries was considered by \cite{2001PhRvD..63h2005C}. Model error effects could also be seen in the parameter estimation analysis of the ``big-dog'' blind injection. In that case, the recovered masses for the compact binary injection were significantly biased (in part) by the fact that different signal models were used for the injection and parameter estimation \cite{2012PhRvD..85h2002A}. This indicates the importance of considering how to incorporate model uncertainties in parameter estimation before the advanced detector era begins. A detailed investigation of parameter estimation on various injections into data from the LIGO/Virgo interferometers and employing a range of different models for the analysis was carried out in~\cite{2013PhRvD..88f2001A}. These results clearly show how the analysis of the same data  using two different models can give mutually inconsistent results.

The recently proposed marginalised likelihood (\cite{Moore&Gair}) attempted to account for the systematic error in the posterior, and hence remove the bias. The approximate likelihood is constructed by including information from a small training set of accurate waveforms computed offline;
\begin{equation}\label{eq:D} {\cal{D}}=\left\{(\vec{\lambda}_{i},\delta h (\vec{\lambda}_{i}))|i=1,2,\ldots,n\right\} \,,\end{equation}
in which $\delta h(\vec{\lambda}) \equiv H(\vec{\lambda})-h(\vec{\lambda})$ denotes the difference between the approximate waveform and the true waveform. GPR assumes that the waveform differences in the training set are a realisation of a Gaussian process with covariance function $k(\vec{\lambda},\vec{\lambda}')$ over the parameter space $\vec{\lambda}$. Different covariance functions may be considered and the \emph{evidence} for the Gaussian process can be maximised with respect to variations in the parameters of the covariance function: this process of optimising the covariance function is called ``training'', and it enables the Gaussian process to ``learn'' the properties of the waveform differences in ${\cal{D}}$. The Gaussian process, once trained, may then be used to interpolate the waveform difference across parameter space. As we are not interested in the actual waveform difference, but rather in its effect on the posterior, the GPR interpolation is used as a prior to analytically marginalise over the unknown waveform difference. The resulting expression for the marginalised likelihood is~\citep{Moore&Gair}
\begin{equation}\label{eq:Lcal} {\cal{L}}(\vec{\lambda})\propto\frac{\exp\left(-\frac{1}{2}\frac{\left<s-H(\vec{\lambda})+\mu(\vec{\lambda}))\big|s-H(\vec{\lambda})+\mu(\vec{\lambda}))\right>}{1+\sigma^{2}(\vec{\lambda})}\right)}{\sqrt{1+\sigma^{2}(\vec{\lambda})}}\,,\end{equation}
where the GPR quantity $\mu(\vec{\lambda})$ is the mean waveform difference and $\sigma^{2}(\vec{\lambda})$ is the error in this GPR estimate;
\begin{eqnarray} &\mu(\vec{\lambda})= k(\vec{\lambda}_{i},\vec{\lambda})\;\textrm{inv}\left(k(\vec{\lambda}_{i},\vec{\lambda}_{j})\right)\;\delta h(\vec{\lambda}_{j})\; ,\label{eq:muGPR}\\
&\sigma^{2}(\vec{\lambda})= k(\vec{\lambda},\vec{\lambda})-k(\vec{\lambda}_{i},\vec{\lambda})\textrm{inv}\left(k(\vec{\lambda}_{i},\vec{\lambda}_{j})\right)k(\vec{\lambda}_{j},\vec{\lambda}) .\label{eq:sigmaGPR} \end{eqnarray}
For more details on the technique of Gaussian process regression see (for example) \cite{MacKay,GPR} and for more details of the marginalised likelihood see \cite{Moore&Gair}.

\section{Analytic calculation of the P-P plot}\label{sec:deriv}
In the limit of high SNR the posterior probability distribution obtained in the analysis of data from a detector will be strongly peaked in the vicinity of the true parameters. Within the vicinity of this peak it is reasonable to expand both the exact and approximate signal models in the usual linear signal approximation (LSA), i.e.\
\begin{eqnarray}h(\vec{\lambda}) &=& h( \vec{\lambda}_0) + \Delta \vec{\lambda}^{a}\partial_{a}h(\vec{\lambda}_0)\,,\nonumber\\
H(\vec{\lambda}) &=& H( \vec{\lambda}_0) + \Delta \vec{\lambda}^{a}\partial_{a}H(\vec{\lambda}_0)\,.\end{eqnarray}
where $\vec{\lambda}_{0}$ denotes the parameter values of the true signal, $\vec{\lambda}$ denotes the parameter values at which we want to evaluate the signal or likelihood and $\Delta \vec{\lambda}=\vec{\lambda}-\vec{\lambda}_{0}$. This LSA is the usual approximation made in the derivation of the Fisher Matrix and the approximation used in the derivation of Eqs.~(\ref{CVsolmain}) and (\ref{secondorder}).

We are interested in predicting the ``sag'' that would be expected in a P-P plot. If we use an approximate waveform model to compute the posterior, then we would expect some bias in the recovered parameters and a sag in the P-P plot - on average the true parameters would be further away from the peak of the posterior than we would expect, and so fewer injections would be recovered at a given significance level.

\subsection{The exact likelihood}\label{subsec:exact}
The exact likelihood, by definition, will give a diagonal unbiased P-P plot. However we will re-derive this obvious result to shed light on the calculations that follow.

The \emph{Exact Likelihood} is given by Eq.~(\ref{eq:Lexact}). The measured data is assumed to consist of a signal with true parameters $\vec{\lambda}_{0}$ and additive Gaussian noise; $s=h(\vec{\lambda}_{0})+n$. In the limit of high SNR,  the difference between two nearby signals in parameter space may be expanded using the LSA,
\begin{eqnarray} L'(\vec{\lambda})&\propto&\exp\left(-\frac{1}{2}\left<n-\Delta \lambda^{a}\partial_{a}h\big|n-\Delta \lambda^{a}\partial_{a}h\right>\right)\,,\\
&=&\exp\left(-\frac{1}{2}\left[\left<n\big|n\right>-2\Delta \lambda^{a}\left<n\big|\partial_{a}h\right>+\Delta\lambda^{a}\Delta\lambda^{b}S_{ab}\right]\right),\nonumber\end{eqnarray}
where the exact Fisher matrix is $S_{ab}=\left<\partial_{a} h|\partial_{b} h\right>$. Since the Fisher matrix is symmetric by construction, we may adopt new coordinates in parameter space $\tilde{\Delta\lambda}^{a}=Q^{a}_{b}\Delta\lambda^{b}$ such that the Fisher matrix in these coordinates becomes diagonal, $S_{ab}=Q^{p}_{a}Q^{q}_{b}\delta_{pq}$. This amounts to rescaling the coordinate axes such that the iso-probability contour, which originally was an $n$-ellipsoid, becomes an $n$-sphere. Derivatives with respect to the new coordinates will be denoted with a tilde, $\partial_{a}h=Q^{b}_{a}\tilde{\partial}_{b}h$. In these new coordinates the likelihood separates to become
\begin{equation} L'(\vec{\lambda})\propto\prod_{x}\exp\left(-\frac{1}{2}\left(\tilde{\Delta\lambda}^{x}-\left<n\big|\tilde{\partial}_{x}h\right>\right)^{2}\right) \,.\end{equation}
In order to exploit the spherical symmetry about the peak in the rescaled parameters we adopt ($n$-dimensional) spherical coordinates centred on the peak; the radial coordinate given by $r^{2}~=~\sum_{x}(\tilde{\Delta\lambda}^{x}~-~{<n|\tilde{\partial}_{x}h>})^{2}$. The significance of the true parameters is given by the volume of the posterior that is ``closer to the peak'', i.e., that has higher posterior weight than the true parameters,
\begin{eqnarray}\label{eq:sig} \textrm{sig}&=&\frac{\int_{0}^{R}\textrm{d}r\;r^{N-1}\exp(-r^{2}/2)}{\int_{0}^{\infty}\textrm{d}r\;r^{N-1}\exp(-r^{2}/2)}\nonumber\\
& = & 1-\frac{\Gamma\left(\frac{N}{2},\frac{R^{2}}{2}\right)}{\Gamma\left(\frac{N}{2}\right)} = 1-\bar{\Gamma}\left(\frac{N}{2},\frac{R^{2}}{2}\right)  \,,\end{eqnarray}
where $\Gamma(x,y)$ is the incomplete Gamma function, 
\begin{equation}
\Gamma(x,y)=\int_y^\infty t^{x-1} {\rm e}^{-t} {\rm d} t \, ,
\end{equation}
$\Gamma(x)=\Gamma(x,0)$ is the complete Gamma function, and $\bar{\Gamma}(x,y)$ is the regularised incomplete gamma function defined via the last equality in Eq.~(\ref{eq:sig}). In Eq.~(\ref{eq:sig}) the assumption has been made that the prior distribution on the parameters may be approximated as a constant across the width of the peak; this is reasonable in the high SNR limit when the posterior is narrow. The quantity $R^{2}$ is given by
\begin{equation}\label{eq:Rsq} R^{2}=\sum_{x}\left<n\big|\tilde{\partial}_{x}h\right>^{2}=\left(S^{-1}\right)^{ab}\left<n\big|\partial_{a}h\right>\left<n\big|\partial_{b}h\right>\, ,\end{equation}
and is distributed as a $\chi^{2}$ random variable with $N=\textrm{dim}(\vec{\lambda})$ degrees of freedom. The inverse regularised incomplete gamma function is defined via $y=\bar{\Gamma}(x,\bar{\Gamma}^{-1}(x,y))$. The quantity on the ordinate axis of a a standard P-P plot is the probability that the true parameters lie within a given significance, $P(\textrm{sig}<X)$. From Eq.~(\ref{eq:Rsq}) it may be seen that this can be rewritten as a cumulative probability of the random variable $R^{2}$;
\begin{equation} P(\textrm{sig}<X)=1-P\left(R^{2}<2\bar{\Gamma}^{-1}\left(\frac{N}{2},1-X\right)\right) \,.\label{eq:numerical}\end{equation}
The cumulative distribution function of the $\chi^{2}$ distribution is the regularised Gamma function, $P(R^{2}<y)=\bar{\Gamma}(N/2,y/2)$. Using this to evaluate Eq.\ \ref{eq:numerical} gives the expected, unbiased diagonal form of the P-P plot for the exact likelihood;
\begin{equation}P(\textrm{sig}<X)=1-(1-X)=X\,.\label{eq:analytic}\end{equation}
This diagonal P-P plot is shown in the dotted black curve in the left-hand panel of Fig.\ \ref{fig:PP}. The fact that the PP plot for the exact likelihood is always diagonal follows from the definition of the likelihood, and this remains true even if the LSA fails. The derivation just presented assumes the LSA in order to make it resemble as closely as possible the upcoming derivation for the approximate likelihood.

\subsection{The approximate likelihood}\label{subsec:approx}
We now move on to the more interesting case when we have biased parameter estimation from using the approximate likelihood. As mentioned in the introduction we expect to obtain a P-P plot that is ``sagging'' below the diagonal indicating the bias. We first treat the simple case where the waveform model depends on just a single parameter, $\vec{\lambda}=\theta$, where the expression for the P-P plot is given in terms of the inverse error function, $\textrm{erf}^{-1}(x)$. A treatment will then be given for the general $N$ dimensional case in which the expression for the P-P plot is given in terms of the MarcumQ function, $Q_{N}(x,y)$, along with a illustration of how this reduces to the 1D result.

The \emph{Approximate Likelihood} is given by Eq.~(\ref{eq:Lapprox}). We assume the approximate model is ``nearly'' correct and use the LSA to expand signals that are nearby in parameter space. As before, denoting the waveform difference by $\delta h (\vec{\lambda})=H(\vec{\lambda})-h(\vec{\lambda})$, we have
\begin{eqnarray} L(\vec{\lambda})&\propto&\exp\left(-\frac{1}{2}\left<n-\delta h(\vec{\lambda}_{0})-\Delta\lambda^{a}\partial_{a}H\big|\ldots\right>\right)\nonumber \\
&=&\exp\left(-\frac{1}{2}\left[\left<n-\delta h(\vec{\lambda}_{0})\Big|\ldots\right>\right.\right.\\ 
&&\;\left.\left.-2\Delta\lambda^{a}\left<n-\delta h(\vec{\lambda}_{0})|\partial_{a}H\right> +\Delta\lambda^{a}\Delta\lambda^{b}\Sigma_{ab} \right]\right) \, ,\nonumber\end{eqnarray}
where the ellipsis in the right hand entry in the inner product denotes a repeat of the left hand entry and the approximate Fisher matrix is $\Sigma_{ab}=\left<\partial_{a} H|\partial_{b} H\right>$. As before coordinates which diagonalise the Fisher matrix $\Sigma_{ab}$ may be adopted, which give the following separated expression for the approximate likelihood,
\begin{equation}\label{eq:Lapproxsep} L(\vec{\lambda})\propto\prod_{x}\exp\left(-\frac{1}{2}\left(\tilde{\Delta\lambda}^{x}-\left<n-\delta h(\vec{\lambda}_{0})\big|\tilde{\partial}_{x}H\right>\right)^{2}\right) \,.\end{equation}

\subsubsection{Example for one dimensional parameter space}
If the waveform depends on only one unknown parameter, $\vec{\lambda}=\lambda$, Eq.~(\ref{eq:Lapproxsep}) becomes
\begin{equation}
L(\theta)  = \frac{1}{\sigma \sqrt{2\pi}} \exp\left[ -\frac{1}{2\sigma^2} \left(\Delta\theta-\mu\right)^2 \right] \, , 
\end{equation}
where
\begin{eqnarray} \frac{1}{\sigma^2} &=& \left<\frac{{\rm d} H}{{\rm d}\lambda}\big|_{\lambda=\lambda_{0}} \Big| \frac{{\rm d} H}{{\rm d}\lambda}\big|_{\lambda=\lambda_{0}} \right>\,, \\
\mu&=&\sigma^2 \left<n - \delta h(\lambda_0) \Big| \frac{{\rm d} H}{{\rm d}\lambda}\big| _{\lambda=\lambda_{0}}\right>\,,\end{eqnarray}
and we have included the correct normalisation of the posterior. The true parameter value is at $\Delta \theta = 0$ and the points with larger posterior weight than the true parameters lie in the range $0 < \Delta \theta < 2 \mu$ when $\mu >0$ or in the range $2 \mu < \Delta \theta < 0$ when $\mu < 0$. The significance at which the true parameters lie is therefore
\begin{equation}
\int_0^{2 \mu} \frac{1}{\sigma \sqrt{2\pi}} \exp\left[ -\frac{1}{2\sigma^2} \left(\Delta\theta-\mu\right)^2 \right] {\rm d}\Delta \theta = {\rm erf} \left( \frac{|\mu|}{\sqrt{2} \sigma} \right)\,, \label{onedsig}
\end{equation}
where 
\begin{equation}
{\rm erf}(z) = \frac{2}{\sqrt{\pi}} \int_0^z {\rm e}^{-t^2} {\rm d} t
\end{equation}
is the usual error function. The quantity $\mu$ defined above depends on the particular realisation of the noise. We want to know the fraction of times, over many realisations of the noise, that the true parameters will lie within a certain significance contour. This is just
\begin{equation}
P({\rm sig} < X) = P\left( \frac{|\mu|}{\sqrt{2} \sigma} < {\rm erf}^{-1} (X) \right) .
\end{equation}
The quantity $\mu/(\sqrt{2} \sigma)$ is distributed as a Gaussian with mean $\tilde{\mu}=\sigma \left<\Delta h(\lambda_0) | {\rm d} H/{\rm d}\lambda\right>/\sqrt{2}$ and variance $1/2$ and so
\begin{eqnarray}P({\rm sig} < X) &=& \frac{1}{2} {\rm erf}\left({\rm erf}^{-1} (X) -\tilde{\mu}\right) \nonumber\\
&&\quad+ \frac{1}{2} {\rm erf}\left({\rm erf}^{-1} (X) +\tilde{\mu} \right) . \label{onedpp}\end{eqnarray}
In the special case where the approximate waveform model and the exact waveform model are the same, we have $\tilde{\mu}=0$ and recover the expected unbiased result from Sec.\ \ref{subsec:exact};
\begin{equation}P({\rm sig} < X) = X\,.\end{equation}

This derivation assumed that $\tilde{\mu}$ was constant, but in practice this will vary from signal to signal. If we denote the probability distribution function for $\tilde{\mu}$ over the astrophysical population by $f(\tilde{\mu})$, the generalisation of Eq.~(\ref{onedpp}) can be seen straightforwardly to be
\begin{eqnarray}
P({\rm sig} < X) &=&\int \left[\frac{1}{2} {\rm erf}\left({\rm erf}^{-1} (X) -\tilde{\mu}\right)  \right. \nonumber\\
&&\quad \left.+ \frac{1}{2} {\rm erf}\left({\rm erf}^{-1} (X) +\tilde{\mu} \right) \right] f(\tilde{\mu}) {\rm d}\tilde{\mu}\,. \label{onedpp:varydh}
\end{eqnarray}

\subsubsection{Parameter space of arbitrary dimension}
We will now generalise the expression for the P-P plot sag in a one-dimensional parameter space, given in Eq.~(\ref{onedpp}), to arbitrary numbers of parameters. Identical manipulations to those performed on the exact likelihood yields the same expression for the significance obtained in Eq.~(\ref{eq:sig}),
\begin{equation}\label{eq:sig2} \textrm{sig}=1-\bar{\Gamma}\left(\frac{N}{2},\frac{R^{2}}{2}\right) \,,\end{equation}
except this time $R^{2}$ is a random variable given by
\begin{eqnarray}\label{eq:Rsqapprox} R^{2}&=&\sum_{x}\left<n-\delta h(\vec{\lambda}_{0})\big|\tilde{\partial}_{x}H\right>^{2} \\
&=&\left(\Sigma^{-1}\right)^{ab}\left<n-\delta h(\vec{\lambda}_{0})\big|\partial_{a}H\right>\left<n-\delta h(\vec{\lambda}_{0})\big|\partial_{b}H\right> \,.\nonumber\end{eqnarray}
If $\delta h(\vec{\lambda})$ is constant across parameter space, $R^{2}$ is now a non-central $\chi^{2}$ random variable with $N$ degrees of freedom and non-centrality parameter
\begin{equation}\label{eq:noncentrality} \Lambda =\left(\Sigma^{-1}\right)^{ab}\left<\delta h(\vec{\lambda}_{0})\big|\partial_{a}H\right>\left<\delta h(\vec{\lambda}_{0})\big|\partial_{b}H\right>\,.\end{equation}
As before the expression for the P-P plot is given in terms of the CDF of the distribution of the random variable $R^{2}$. The CDF of the non-central $\chi^{2}$ distribution is the Marcum-Q function, $P(R^{2}~<~y)=Q_{N/2}(\sqrt{\Lambda},\sqrt{y})$, 
\begin{eqnarray} P(\textrm{sig}<X)&=&1-P\left(R^{2}<2\bar{\Gamma}^{-1}\left(\frac{N}{2},1-X\right)\right) \label{eq:numerical2}\\
&=&1-Q_{\frac{N}{2}}\left(\sqrt{\Lambda},\sqrt{2\bar{\Gamma}^{-1}\left(\frac{N}{2},1-X\right)}\right)\,.\label{eq:analytic2}\end{eqnarray}
This is an analytic approximation to the P-P plot in the LSA and in the case of a constant waveform difference over parameter space; this function is plotted as a dotted black line in Fig.\ \ref{fig:PP}. In this case the P-P plot always sags below the diagonal indicating biased parameter recovery.

If $\delta h(\vec{\lambda})$ is not constant over parameter space, the generalisation of this result takes the same form as Eq.~(\ref{onedpp:varydh}), but with the term in square brackets replaced by Eq.~(\ref{eq:analytic2}) and with $f(\tilde{\mu})$ replaced by the corresponding probability distribution function for $\Lambda$. For example, in the case that $\delta h(\vec{\lambda})$ is distributed at different times and at different points in parameter space as an uncorrelated, zero-mean Gaussian with variance in each component of $\epsilon^{2}$ (i.e.\ $\delta h(\vec{\lambda}_{0})\sim{\cal{N}}(0,\epsilon^{2})$) then the quantities $\langle \delta h(\vec{\lambda}_0) | \partial_a H\rangle$ are distributed as $N(0, \Sigma)$ and we see that $\Lambda$ is distributed as $\epsilon^2$ times a $\chi^2$ distribution with $N$ degrees of freedom with probability distribution function
\begin{equation}
f(\Lambda) = \frac{1}{2^{\frac{N}{2}} \Gamma(N/2) \epsilon^N} \Lambda^{\frac{N}{2}-1}{\rm e}^{-\frac{\Lambda}{2 \epsilon^2}} .
\end{equation}
Writing $x_u^2 = 2 \bar{\Gamma}^{-1}\left(\frac{N}{2},1-X\right)$ we must evaluate
\begin{eqnarray}
&&\hspace{-0.25in}P(\textrm{sig}<X) \nonumber \\
&=& \int_0^\infty\left[ \frac{\Lambda^{\frac{N}{2}-1}{\rm e}^{-\frac{\Lambda}{2 \epsilon^2}} }{2^{\frac{N}{2}}  \Gamma\left(\frac{N}{2}\right) \epsilon^N}  \right. \nonumber \\
&& \qquad\left. \int_0^{x_u} x\left(\frac{x}{\sqrt{\Lambda}}\right)^{\frac{N}{2}-1} {\rm e}^{-\frac{1}{2} (x^2 + \Lambda)} I_{\frac{N}{2}-1}(\sqrt\Lambda x) {\rm d x} \right] {\rm d \Lambda} \nonumber \\
&=& \frac{1}{(2\epsilon)^{\frac{N}{2}-1} \Gamma\left(\frac{N}{2}\right)} \int_0^{x_u} \left[ x^{\frac{N}{2}} {\rm e}^{-\frac{x^2}{2}} \right. \nonumber \\
&& \qquad \qquad\left.\int_0^\infty y^{\frac{N}{2}} {\rm e}^{-\frac{1}{2}(1+\epsilon^2)y^2} I_{\frac{N}{2}-1}(\epsilon x y) {\rm d}y \right] {\rm d}x \nonumber \\
&=& \frac{1}{(2\epsilon)^{\frac{N}{2}-1} \Gamma\left(\frac{N}{2}\right) (1+\epsilon^2)^{\frac{1}{2}+\frac{N}{4}}} \int_0^{x_u} \left[ x^{\frac{N}{2}} {\rm e}^{-\frac{x^2}{2}} \right. \nonumber \\
&& \qquad \qquad\left.\int_0^\infty \tilde{y}^{\frac{N}{2}} {\rm e}^{-\frac{1}{2}\tilde{y}^2} I_{\frac{N}{2}-1}\left(\frac{\epsilon}{\sqrt{1+\epsilon^2}} x \tilde{y}\right) {\rm d}\tilde{y} \right] {\rm d}x \nonumber \\
&=& \frac{1}{2^{\frac{N}{2}-1} \Gamma\left(\frac{N}{2}\right) (1+\epsilon^2)^{\frac{N}{2}}} \int_0^{x_u}  x^{N-1} {\rm e}^{-\frac{x^2}{2(1+\epsilon^2)}} {\rm d} x \nonumber \\
&=& \frac{1}{2^{\frac{N}{2}} \Gamma\left(\frac{N}{2}\right)}  \int_0^{\frac{x_u^2}{1+\epsilon^2}}  u^{\frac{N}{2}-1} {\rm e}^{-\frac{u}{2}} {\rm d} u \nonumber \\
&=&1-\bar{\Gamma}\left(\frac{N}{2},\frac{\bar{\Gamma}^{-1}\left(\frac{N}{2},1-X\right)}{1+\epsilon^{2}}\right) \, ,\label{eq:avoverGaussian}
\end{eqnarray}
where the second line follows by a change of variable and a change in the order of integration, the fourth line follows from the fact that $Q_m(a,0)=1$ and the final lines follow from another change of variable. We have also made use of the integral expression for the Marcum-Q function given below. This result can also be obtained directly by noticing that the random variable $R^{2}$, which depends on both $n$ and $\delta h(\vec{\lambda})$, is distributed as $1+\epsilon^{2}$ times a $\chi^{2}$ random variable with $N$ degrees of freedom. The analytic expression for the P-P plot is therefore very similar to the case of the exact likelihood, but with the argument of the regularised Gamma function scaled appropriately to give the same result as in the final line of Eq.(\ref{eq:avoverGaussian}). This P-P plot also exhibits a sag below the diagonal; see the orange dotted curve in Fig.\ \ref{fig:PP}. 

The $N$ dimensional result in Eq.~(\ref{eq:analytic2}) can be shown to reduce to the 1 dimensional result in Eq.~(\ref{onedpp}) using the standard properties of the Marcum-Q function. The Marcum-Q function is defined by the integral
\begin{eqnarray}Q_m(a,b) &=& \int_b^\infty x \left(\frac{x}{a}\right)^{m-1} \exp\left[-\frac{1}{2} (x^2+a^2) \right] I_{m-1}(ax) {\rm d}x \nonumber \\ 
&=& \exp\left[-\frac{1}{2} (a^2+b^2) \right] \sum_{k={1-m}}^\infty \left(\frac{a}{b}\right)^k I_k (ab) \, ,\label{marcumQ}\end{eqnarray}
in which $I_n(x)$ is the modified Bessel function of the first kind.  For $N=1$, the Marcum-Q function, Eq.~(\ref{marcumQ}), can also be simplified
\begin{eqnarray}
Q_{\frac{1}{2}}(a,b) &=& \sqrt{a} \int_b^\infty \sqrt{x} \exp\left[-\frac{(x^2+a^2)}{2}\right] I_{-1/2}(ax) {\rm d}x \nonumber\\ &=& \sqrt{\frac{1}{2\pi}} \int_b^\infty \left( \exp\left[-\frac{(x+a)^2}{2}\right] + \exp\left[-\frac{(x-a)^2}{2}\right] \right) {\rm d}x \nonumber \\
&=& 1 - \frac{1}{2} \left( {\rm erf} \left(\frac{b-a}{\sqrt{2}}\right) + {\rm erf} \left(\frac{b+a}{\sqrt{2}}\right) \right) \, ,
\end{eqnarray}
which follows from $I_{-\frac{1}{2}}(x) = \sqrt{2/\pi} \cosh(x) / \sqrt{x}$. When $N=1$ the regularised Gamma function becomes
\begin{eqnarray}
\frac{\Gamma\left(1/2, R^2/2 \right)}{\Gamma\left(1/2\right)} &=& \frac{ \int_{R^2/2}^\infty {\rm e}^{-t}/\sqrt{t} \;\;{\rm d} t}{ \int_0^\infty {\rm e}^{-t}/\sqrt{t} \;\;{\rm d} t} \nonumber \\
&=& \frac{ \int_{R/\sqrt{2}}^\infty {\rm e}^{-u^2} \;\;{\rm d} u}{ \int_0^\infty {\rm e}^{-u^2} \;\;{\rm d} u} \nonumber \\
&=& 1-{\rm erf}(R/\sqrt{2}) \, .
\end{eqnarray}
Eq.~(\ref{eq:numerical2}) therefore becomes
\begin{eqnarray}
P({\rm sig} < X) &=& \frac{1}{2} \left( \rm{erf} \left( {\rm erf}^{-1}(X) - \sqrt{\frac{\Lambda}{2}}\right) \right.\nonumber\\
&&\quad\;\left.+ \rm{erf} \left( {\rm erf}^{-1}(X) + \sqrt{\frac{\Lambda}{2}}\right) \right) \, ,
\end{eqnarray}
as we can identify $\tilde{\mu} = \Lambda/2$, we recover Eq.~(\ref{onedpp}) as expected.

\subsection{The marginalised likelihood}\label{subsec:marg}
\begin{figure*}[t]
 \centering
 \includegraphics[trim=0cm 0cm 0cm 0cm, width=0.98\textwidth]{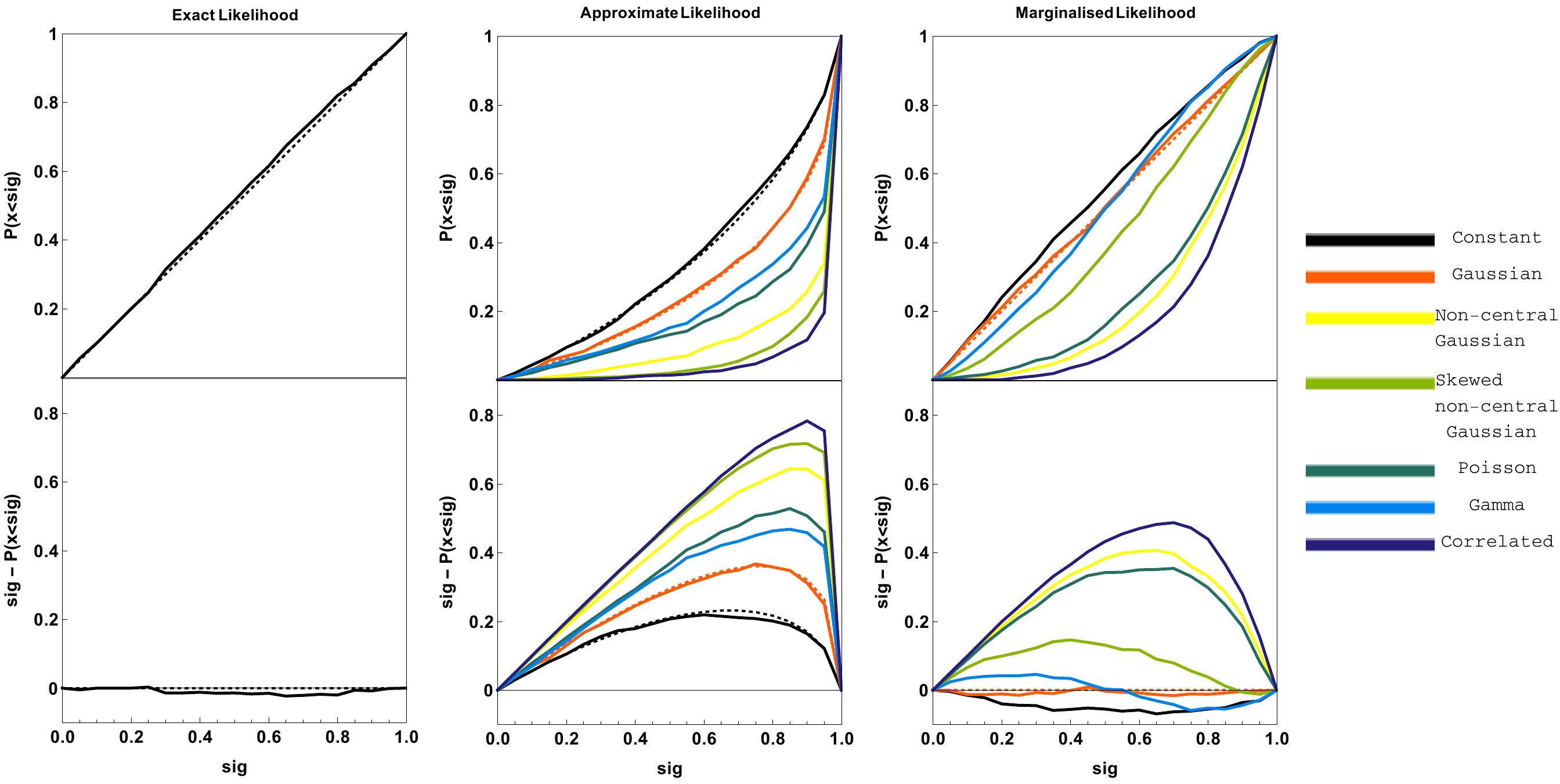}
 \caption{P-P plots for parameter estimation using the three likelihoods $L'(\vec{\lambda})$, $L(\vec{\lambda})$, and ${\cal{L}}(\vec{\lambda})$ shown in the three columns respectively. In each column the top panel shows a P-P plot whilst the bottom panel shows the ``sag''; i.e. the difference between the ideal diagonal line and the actual P-P plot. In each panel curves drawn as dotted lines correspond to analytic results whilst solid curves are numerical results. The left-hand column shows ideal, unbiased parameter recovery for the exact likelihood. In the centre and right-hand columns different colour curves correspond to different distributions of $\small<\delta h(\lambda_{0})|\tilde{\partial}_{x}H\small>$. The curves in black are for a constant distribution giving a non-centrality $\Lambda=2$ ($\Lambda$ defined in Eq.~(\ref{eq:noncentrality})). The curves in orange are for a zero mean Gaussian distribution with variance $1$. The curves in yellow are for a non-central Gaussian distribution with mean $4/3$ and variance $1$. The curves in light-green are for a non-central, skewed normal distribution\footnote{The PDF of a skew Gaussian distribution with location parameter $\mu$, scale parameter $\sigma$ and skew parameter $\alpha$ is given by $\left[1+\textrm{erf}(\alpha (x-\mu)/{\sqrt{2}\sigma})\right]\exp ({-(x-\mu)^{2}}/{2\sigma^{2}})$} with location parameter $1$, scale parameter $1$, and skew parameter $1$. The curves in dark-green are for a Poisson distribution with mean and variance $1$. The curves in blue are for a Gamma distribution with shape parameter $1$ and scale parameter $1$. And finally, the curves in purple are for a correlated random walk distribution with mean Gaussian step size $1$. In all cases the number of parameter dimensions is $N=4$, and the number of points used for the numerical simulations was $n=10^{3}$. The left-hand panel clearly shows the exact likelihood does not suffer from any bias, as expected. The centre panel shows that in all cases the approximate likelihood suffers from a bias. The right-hand column shows that in all cases the marginalised likelihood reduces the bias relative to the approximate likelihood. In the ideal case (shown in orange) of a zero mean Gaussian distribution for $\small<\delta h(\lambda_{0})|\tilde{\partial}_{x}H\small>$ the bias is completely removed.}
 \label{fig:PP}
\end{figure*}

The \emph{Marginalised Likelihood} is given by Eq.~(\ref{eq:Lcal}), as before this may be expanded in the LSA. In the high SNR limit the posterior is narrow compared to the length scale over which the waveform changes. The waveform difference changes over the same length scale as the waveform. The quantity $\sigma^{2}(\vec{\lambda})$ also changes over this length scale, as it is ``learnt'' by the GP in the procedure of maximising the evidence. Therefore in the high SNR limit $\sigma^{2}(\vec{\lambda})$ may be approximated as a constant. As before coordinates which diagonalise the Fisher matrix may be adopted, which give the following separated expression for the approximate likelihood,
\begin{widetext}
\begin{equation}\label{eq:margLseperated} {\cal{L}}(\vec{\lambda})\propto\prod_{x}\exp\left(-\frac{1}{2}\frac{\left(\tilde{\Delta\lambda}^{x}-\left<n+\mu(\vec{\lambda}_{0})-\delta h(\vec{\lambda}_{0})\big|\tilde{\partial}_{x}(H-\mu)\right>\right)^{2}}{1+\sigma^{2}}\right) \,.\end{equation}
\end{widetext}
The waveform difference is assumed to be a small quantity, therefore in Eq.~(\ref{eq:margLseperated}) the derivative $\tilde{\partial}_{x}(H-\mu)$ may be replaced by $\tilde{\partial}_{x}(H)$, as the difference is the product of small quantities. Identical manipulations to those performed on the exact and approximate likelihoods give the same expression for the significance obtained in Eqs.~(\ref{eq:sig}) and (\ref{eq:sig2}),
\begin{equation} \textrm{sig}=1-\bar{\Gamma}\left(\frac{N}{2},\frac{R^{2}}{2}\right) \,,\end{equation}
except this time the random variable $R^{2}$ is given by
\begin{widetext}
\begin{eqnarray}\label{eq:Rsqmarg} R^{2}&=&\frac{1}{1+\sigma^2} \sum_{x}\left<n+\mu(\vec{\lambda}_{0})-\delta h (\vec{\lambda}_{0})\big| \tilde{\partial}_{x}H\right>^{2}\,,\\
&=&\frac{1}{1+\sigma^{2}} \left(\Sigma^{-1}\right)^{ab}\left<n+\mu(\vec{\lambda}_{0})-\delta h(\vec{\lambda}_{0})\big|\partial_{a}H\right>\left<n+\mu(\vec{\lambda}_{0})-\delta h(\vec{\lambda}_{0})\big|\partial_{b}H\right> \,.\end{eqnarray}
\end{widetext}
The GPR technique assumes that the $\delta h(\vec{\lambda})$ are distributed as a Gaussian process across parameter space, with zero mean and a covariance estimated from a training set and any prior knowledge. If this assumption is in fact true, and the covariance has been correctly estimated, then the quantity $\mu(\vec{\lambda}_{0})-\delta h(\vec{\lambda}_{0})$ is distributed as a zero mean Gaussian with variance $\sigma^{2}$. In this case (perhaps unsurprisingly) the marginalised likelihood completely fixes the sag. The new $R^{2}$ random variable is distributed as 
a $\chi^{2}$ random variable with $N$ degrees of freedom and using the regularised Gamma function as the CDF of this distribution we recover the diagonal P-P plot;  
\begin{eqnarray} P(\textrm{sig}<X)&=&1-P\left(R^{2}<2\bar{\Gamma}^{-1}\left(\frac{N}{2},1-X\right)\right), \label{eq:numerical3}\\
&=&X\,.\label{eq:analytic3} \end{eqnarray}
This case is shown, both analytically and numerically, in orange in Fig.\ \ref{fig:PP}.

More interesting is the behaviour in the realistic case when $\delta h(\vec{\lambda})$ is not distributed exactly as the GPR has predicted. This case is more complicated because the different components that make up the $R^{2}$ random variable are no longer independent random variables and a simple expression for the distribution of $\delta h (\vec{\lambda})$ cannot be found. In particular, from Eq.~(\ref{eq:Rsqmarg}) it can be seen that $R^{2}$ is the sum of the squares of a noise term, $\small<n|\tilde{\partial}_{x}H\small>$, a GPR term, $\small<\mu(\vec{\lambda}_{0})|\tilde{\partial}_{x}H\small>$, and (minus) a physical term, $\small<\delta h(\vec{\lambda}_{0})|\tilde{\partial}_{x}H\small>$. In particular the GPR and physical terms are now related because the expression for $\mu(\vec{\lambda}_{0})$ in Eq.~(\ref{eq:muGPR}) is a linear combination of the realisations of $\delta h(\vec{\lambda})$ in the training set, ${\cal{D}}$. The sag will still be given by the analogue of Eq.~(\ref{onedpp:varydh}), but this integral will not in general be analytically tractable. Instead, we will consider such cases numerically in Sec.\ \ref{sec:numerics}.

As we have seen, in the particular case considered above where the waveform difference is distributed as assumed by the GPR, the marginalised likelihood completely removes the systematic bias present in the standard, approximate, likelihood. In addition, as we will see in Sec.\ \ref{sec:numerics}, even in unfavourable situations the marginalised likelihood is often able to remove significant portions of the bias. We conclude this section with a discussion of why it is expected that the bias in parameter estimates obtained using the marginalised likelihood will usually be less than those obtained using the standard likelihood.

From Eqs.~(\ref{eq:Rsqapprox}) and (\ref{eq:Rsqmarg}) it can be seen that the condition for the marginalised likelihood to yield \emph{more} biased parameter estimates than the approximate likelihood, {\it for a particular event}, is $R^2_{\textrm{approx}}<R^2_{\textrm{GPR}}/(1+\sigma^{2}(\vec{\lambda}_{0}))$, where
\begin{eqnarray} R^2_{\textrm{approx}}&=&\sum_{x}\left<n-\delta h (\vec{\lambda}_{0})\big|\tilde{\partial}_{x}H\right>^2 \nonumber\\
R^2_{\textrm{GPR}}&=&\sum_{x}\left<n-\delta h (\vec{\lambda}_{0})+\mu (\vec{\lambda}_{0})\big|\tilde{\partial}_{x}H\right>^2\,.\end{eqnarray}
These terms both involve a projection onto the space spanned by the derivatives, $\tilde{\partial}_{x}H$, at the point $\vec{\lambda}_{0}$.
Since these ``tilde'' derivatives were constructed to be an orthonormal basis, the condition for the marginalised likelihood to give worse parameter estimates than the approximate likelihood can therefore be written as
\begin{equation} \left|n-\delta h (\vec{\lambda}_{0})\right|_{\cal D}^{2} < \frac{\left| n-\delta h (\vec{\lambda}_{0})+\mu (\vec{\lambda}_{0})\right|_{\cal D}^{2}}{1+\sigma^{2}(\vec{\lambda}_{0})}\,
\label{eq:worsesagcond}\end{equation}
where the modulus is taken with respect to the function inner product in Eq.~(\ref{eq:inner}), projected into the space, ${\cal D}$, spanned by the derivatives. For this to be satisfied, it would be necessary not only for the interpolation to have the wrong sign when expressed in the basis $\tilde{\partial}_{x}H$ (i.e. $0>\sum_{x}\left<\right.h(\vec{\lambda}_{0})|\tilde{\partial}_{x} H\left.\right> \left<\right.\mu(\vec{\lambda}_{0})|\tilde{\partial}_{x} H\left.\right>$), but also for it to be large enough in magnitude to overcome the GPR uncertainty $\sigma^{2}(\vec{\lambda}_{0})$ in the denominator. Moreover, this is just for one particular realisation of the noise and true waveform parameters. We are really interested in the sag that arises when considering a population of events. In that case, we would need Eq.~(\ref{eq:worsesagcond}) to be true in some average sense and so the interpolation would have to have the wrong sign and be too large for the majority of choices of waveform parameters. Although this is technically possible, it is clear that any reasonable interpolation algorithm with decent coverage of the parameter space in the training set and a reasonable covariance function should violate the above bound on average and therefore yield better parameter estimates on average and show a smaller sag in the P-P plot than the approximate likelihood.

\section{Numerical calculation of the P-P plot}\label{sec:numerics}
In all of the above calculations the expression for the P-P plot was written in terms of the CDF of the distribution of the $R^{2}$ random variable. This random variable is written in terms of a signal inner product of the model derivatives, it therefore depends both on the properties of the GW source and of the GW detector. By expressing our results in terms of $R^{2}$ we ensure that they remain valid for any detector and any source (assuming the LSA holds). In the cases considered above where analytic expression for the P-P plots could be found these can also be verified numerically by drawing $n$ values of $R^{2}$ from the relevant distribution and numerically estimating the CDF. In cases where an analytic expression for the P-P plot can not be found the same procedure can be used to investigate the P-P plot numerically.

First consider the unbiased, diagonal P-P plot obtained for the exact likelihood. The analytic expression for this P-P plot is given in Eq.~(\ref{eq:analytic}). A numerical validation of this result may be performed by drawing random realisations of the $R^{2}$ value in Eq.~(\ref{eq:Rsq}). It can be seen that $R^{2}$ is the sum of the squares of $N$ standard Gaussian random variables $\small< n| \tilde{\partial}_{x}H \small>$; i.e. a $\chi^{2}$ distribution with $N$ degrees of freedom. We drew $n$ realisations of $R^{2}$ from this distribution, numerically estimated the CDF and plotted the P-P plot using Eq.~(\ref{eq:numerical}). The results for $n=10^{3}$ and $N=4$ are shown in the left panels of Fig.\ \ref{fig:PP} (analytic results shown as a dotted line, numerical results as a solid line). Within the scale of fluctuations the numerical results agree well with the analytic results. The bottom left panel of the same figure shows the sag of the P-P plot below the diagonal, i.e. $\textrm{sig}-P(x<\textrm{sig})$. The values $n=10^{3}$ and $N=4$ will also be used for all subsequent numerical calculations in this section.

P-P plots for the approximate likelihood are shown in the centre panels of Fig.~\ref{fig:PP} for a variety of different distributions of the waveform difference projected into the model derivatives; $\small<\delta h(\lambda_{0})|\tilde{\partial}_{x}H\small>$. In the case of a constant distribution, or a zero-mean Gaussian distribution the analytic expressions in Eqs.~(\ref{eq:analytic2}) and (\ref{eq:avoverGaussian}) respectively are shown as dotted lines. For the numerical calculations the procedure followed was first to specify the distribution for the $\small<\delta h(\lambda_{0})|\tilde{\partial}_{x}H\small>$ random variables (for example the black curves show results when this is a constant). The quantity $R^{2}$ was then calculated using Eq.~(\ref{eq:Rsqapprox}) by drawing a random value from this distribution and a random value for $\small<n|\tilde{\partial}_{x}H\small>$ from a standard Gaussian distribution. The $R^{2}$ variable was calculated $n$ times, the CDF of this variable estimated, and the P-P plot calculated from Eq.~(\ref{eq:numerical2}). Different colours in Fig.~\ref{fig:PP} indicate different distributions for $\small<\delta h(\lambda_{0})|\tilde{\partial}_{x}H\small>$, the specification of these distributions are given in the figure caption. 

\begin{table}[h]
  \begin{tabular}{ l | c  c }
 & Approximate & Marginalised \\
\hline
Constant & 0.158 & -0.044 \\
Gaussian & 0.237 & 0.000 \\
non-central Gaussian & 0.385 & 0.263 \\
Skew non-central Gaussian & 0.426 & 0.079 \\
Poisson & 0.317 & 0.235 \\
Gamma & 0.293 & -0.001 \\
Correlated & 0.441 & 0.308 \\
  \end{tabular}
  \caption{Table of the total integrated biases for the curves shown in Fig.~\ref{fig:PP}. The integrated bias is defined as the total area in the sag, i.e. $\int_{0}^{1}\textrm{d}(\textrm{sig})\;(\textrm{sig}-P(x<\textrm{sig}))$. \label{tab:1} }
\end{table}

P-P plots for the marginalised likelihood are shown in the right panels of Fig.~\ref{fig:PP} for a variety of different distributions of $\small<\delta h(\lambda_{0})|\tilde{\partial}_{x}H\small>$. In the case of a zero-mean Gaussian the analytic expressions in Eq.~(\ref{eq:analytic3}) is shown as a dotted line. For the numerical calculations it is necessary to construct a training set for the GPR interpolation. Instead of using GPR to interpolate the waveform differences, $\delta h (\vec{\lambda})$, it is simpler for our present purpose to instead interpolate the projections of the waveform differences onto the waveform derivatives, i.e., $\small<\delta h(\vec{\lambda})|\tilde{\partial}_{x}H\small>$, as these are what appear in Eq.~(\ref{eq:Rsqmarg}), . The training set was taken to consist of points at $\lambda=1,2,\ldots,20$ and the actual experimental realisation at a value $\lambda_{0}=21$. For the majority of distributions (constant, Gaussian, non-central Gaussian, skew non-central Gaussian, Poisson, and Gamma distributions) shown in Fig.~\ref{fig:PP} the random variables in the training set were drawn independently and interpolated using an uncorrelated Gaussian process, i.e. $K_{ij}=\sigma_{f}\delta_{ij}$. The $R^{2}$ value was calculated from Eq.~\ref{eq:Rsqmarg} (with $\mu=0$ from Eq.~(\ref{eq:muGPR}), because of the assumption of an uncorrelated process), the CDF estimated and the P-P plot calculated from Eq.~(\ref{eq:numerical3}). 

The assumption of an uncorrelated Gaussian process is a conservative assumption. In the absence of correlations the Gaussian process regression assumes a ``worst-case'' scenario and returns a mean waveform difference of zero (see Eq.~(\ref{eq:muGPR})). If correlations were present then the GPR would return a non-zero estimate for $\mu$ and shift the position of the posterior peak into better agreement with the true value, thus improving the P-P plot. To investigate the effect of correlations the final numerical calculation (labelled as ``correlated'' in Fig.~\ref{fig:PP}) was performed using a random walk distribution. The values of  $\small<\delta h(\vec{\lambda})|\tilde{\partial}_{x}H\small>$ at the points $\lambda=1,2,\ldots,21$ were taken to be a realisation of a random walk with Gaussian step width $a=1/3$. The first 20 of these values were taken as the training set and used to extrapolate the final value. For the GPR interpolation a squared exponential covariance function $k(x,y)=\exp((-1/2)(x-y)^{2})$ was used. The squared exponential covariance function is not able to accurately capture the covariance of the random walk distribution, so this again represents a conservative choice to examine how the marginalised likelihood performs in the presence of un-modelled correlations. However, even in this unfavourable case the marginalised likelihood still significantly reduces the bias in the P-P plot.

The purpose of considering such a wide variety of different distributions for the waveform difference is to test whether the marginalised likelihood is robust against different types of errors in the waveform models, which are not correctly modelled by the Gaussian process. For example, the marginalised likelihood assumes the waveform difference is a zero mean Gaussian process across parameter space, therefore it is perhaps not surprising that it performs well in the case of a zero mean Gaussian distribution. However the list of distributions used here also test the robustness of the method against non-central distributions (e.g non-central Gaussian), skewed distributions (e.g. skewed Gaussian), one-sided and non-Gaussian distributions (e.g. Poisson or Gamma distributions), and the presence of un-modelled correlations in the waveform difference (the random walk distribution).

By comparing the curves of the same colour between the centre and right-hand panels of Fig.\ \ref{fig:PP} it can be seen that in all cases the P-P plot for the marginalised likelihood exhibits less of a bias, i.e.,\ less of a ``sag'', than the approximate likelihood. In the ideal case where the distribution of the waveform differences is precisely that assumed by the GPR, a diagonal, unbiased P-P plot is recovered; however the bias is also almost completely removed for several of the other distributions considered. In all cases a significant improvement in performance can be seen when using the marginalised likelihood in place of the standard approximate likelihood. These results are summarised in Table~\ref{tab:1}, which lists the \emph{total bias} (defined as the area between the sagging curve and the ideal diagonal) for all the curves shown in Fig.~\ref{fig:PP}.

\section{Discussion}\label{sec:conclusions}
The P-P plot provides a way to quantify the bias that results when using inaccurate models to perform GW parameter estimation. For individual sources the systematic error in the parameters is independent of the SNR, whilst the random errors scale as $1/\textrm{SNR}$, and hence the bias is most significant for the loudest sources. Even in cases where, for each individual source, the systematic error is small compared to the random error, the bias can still be significant when observing populations of sources, since the statistical error in a parameter estimated from combining a population of sources reduces as $1/\sqrt{N}$ as more sources are added, while the systematic errors remain fixed.

In this paper several analytic expressions have been obtained that predict the sag of the P-P plots that results from different distributions of the model error. These results have been derived within the linear signal approximation, and are valid to ${\cal{O}}(1/\textrm{SNR})$. These analytic expressions for the P-P plots may be viewed in the same spirit as Fisher matrix estimates for the random errors, or Cutler and Vallisneri's \cite{CV} expression for the systematic error in a single measurement. This latter result has also here been generalised (in Appendix~\ref{appendix}) to include terms of ${\cal{O}}(1/\textrm{SNR}^{2})$.

It is now well established that model errors will present significant problems for a range of GW sources. The authors recently proposed a novel method for tackling this problem; using a modified likelihood constructed using Gaussian process regression on a training set of accurate waveforms. In this paper the performance of this marginalised likelihood was examined by comparing the P-P plots (obtained both analytically and numerically) with those obtained from the standard likelihood. In particular, it was found that in favourable conditions the marginalised likelihood was able to completely remove the parameter estimation bias. More importantly, it was found that the marginalised likelihood was robust against a wide range of un-modelled features in the distribution of waveform differences, and in all cases considered outperformed the standard likelihood. These results provide further illustration of the need to account for model uncertainties (using GPR or other techniques) when drawing inferences from near future GW observations.

\bibliography{bibliography}

\begin{appendix}
\section{Systematic bias due to waveform errors}\label{appendix}
We assume that an approximate model $H(\vec{\lambda})$ is used to recover the parameters of a gravitational wave signal that is described by the true model $h(\vec{\lambda})$ with parameters $\vec{\lambda}_0$. The best fit parameters of the approximate model are $\vec{\lambda}_{\rm bf} = \vec{\lambda}_0 + \Delta \vec{\lambda}$. These parameters minimise the squared distance between the true and approximate model spaces, 
\begin{equation} \left<\delta h(\vec{\lambda}_{0})+H(\vec{\lambda}_{\textrm{bf}})-H(\vec{\lambda}_{0})\big|\delta h(\vec{\lambda}_{0})+H(\vec{\lambda}_{\textrm{bf}})-H(\vec{\lambda}_{0})\right> \,\end{equation}
where $\delta h (\vec{\lambda}_{0})=H(\vec{\lambda}_{0})-h(\vec{\lambda}_{0})$ (note the different sign convention from \cite{CV}). Differentiating with respect to each of the parameters in turn, we find that the best-fit parameters must satisfy the equations
\begin{equation}\langle \delta h(\vec{\lambda}_{0}) +H(\vec{\lambda}_{\textrm{bf}})-H(\vec{\lambda}_{0})| \partial_{a} \left(H(\vec{\lambda}_{\textrm{bf}})-H(\vec{\lambda}_{0})\right) \rangle =0\,. \label{maxeq}\end{equation}
We use the notation $\partial_{a} x \equiv \partial x/\partial \lambda^{a}$ and subsequently will use $\partial_{ab} x\equiv \partial^2 x/\partial \lambda^{a} \partial\lambda^{b}$. If we now assume that the approximation is good, we can write $\delta h (\vec{\lambda})\sim {\cal{O}}(\epsilon)$, a small parameter, and $\vec{\lambda}_{\textrm{bf}} = \vec{\lambda}_{0} + \vec{\Delta \lambda}$ with $\Delta\lambda^{i} \sim {\cal{O}}(\epsilon) \,\, \forall i$. We can then expand the difference between the approximate waveforms as a Taylor series
\begin{eqnarray}
H(\vec{\lambda}_{\textrm{bf}})-H(\vec{\lambda}_{0})&=&\partial_{a}H(\vec{\lambda}_{0})\Delta\lambda^{a}+ \nonumber\\
&&\frac{1}{2}\partial_{ab} H(\vec{\lambda}_0) \Delta \lambda^{a} \Delta \lambda^{b} +\cdots \,.\label{eq:TaylorExpand}
\end{eqnarray}
Eq.~(\ref{maxeq}) becomes
\begin{widetext}
\begin{eqnarray}
\left<\delta h(\vec{\lambda}_{0})+\partial_{b}H(\vec{\lambda}_{0})\Delta\lambda^{b}+\frac{1}{2}\partial_{bc} H(\vec{\lambda}_0) \Delta \lambda^{b} \Delta \lambda^{c}\big|\partial_{a}H(\vec{\lambda}_{0})+\partial_{ad} H(\vec{\lambda}_0) \Delta \lambda^{d}\right>=0\,.
\end{eqnarray}
\end{widetext}
where all derivatives are now evaluated at $\vec{\lambda}_0$. Keeping only terms of order $\epsilon$ we find the Cutler and Vallisneri result
\begin{equation}
\Delta\lambda_{1}^{a} = - \left( \Sigma^{-1} \right)^{ab} \langle \delta h(\vec{\lambda}_{0}) | \partial_{b} H(\vec{\lambda}_{0}) \rangle \label{CVsol}
\end{equation}
where $\Sigma^{ij} \equiv \langle \partial_a H(\vec{\lambda}_{0}) | \partial_b H(\vec{\lambda}_{0}) \rangle$ is the Fisher Matrix.

We now extend to the next order in $\epsilon$ by writing $\Delta \lambda^{a} = \Delta\lambda_{1}^{a} + \Delta\lambda_{2}^{a}$, where $\Delta\lambda_1^i$ is the previous solution, Eq.~(\ref{CVsol}). Keeping terms to ${\cal{O}}(\epsilon^2)$ we obtain
\begin{widetext}
\begin{equation}
\Delta\lambda_{2}^{a} = -\left( \Sigma^{-1} \right)^{ab} \left[ \langle \delta h (\vec{\lambda}_{0}) | \partial_{ab} H(\vec{\lambda}_{0}) \rangle \Delta \lambda_{1}^{b} + \langle \partial_{ac} H(\vec{\lambda}_{0}) | \partial_{b} H(\vec{\lambda}_{0}) \rangle \Delta \lambda_{1}^{b} \Delta \lambda_{1}^{c} + \frac{1}{2} \langle \partial_a H(\vec{\lambda}_{0}) | \partial_{ab} H(\vec{\lambda}_{0}) \rangle \Delta \lambda_{1}^{b} \Delta \lambda_{1}^{c} \right]. \label{secondorder}
\end{equation}
\end{widetext}
A suitable validity criterion for the Cutler and Vallisneri formula, (\ref{CVsol}), is
\begin{equation}
\mbox{max}_{a} \left\{|\Delta\lambda^{a}_{2}/\Delta\lambda^{a}_{1}|\right\} \ll 1.
\end{equation}
We note also that Eq.~(\ref{secondorder}) provides an improved estimate of the systematic bias and that we can readily extend this method to higher order in $\epsilon$ by including further terms in the expansion in Eq.~(\ref{eq:TaylorExpand}).
\end{appendix}

\end{document}